\documentclass[12pt,preprint]{aastex}
\usepackage{url}
\usepackage{float}
\usepackage{multirow}
\usepackage[figuresright]{rotating}
\usepackage{amssymb}
\usepackage{amsmath}
\usepackage{epstopdf}
\usepackage{natbib}
\usepackage{xcolor}
\usepackage[colorlinks, citecolor=blue]{hyperref}
\usepackage[all]{hypcap}
\usepackage[flushleft]{threeparttable}
\usepackage{lscape}
\usepackage{longtable}
\usepackage{url}
\usepackage{epsfig}
\usepackage{lineno}
\shorttitle{Evidence for gravitational lensing of GRB 200716C}
\shortauthors{Yang et al.}
\slugcomment{}

\begin{document}

\title{Evidence for gravitational lensing of GRB 200716C}
\author{Xing Yang\altaffilmark{1}, Hou-Jun L\"{u}\altaffilmark{1}, Hao-Yu Yuan\altaffilmark{1},
Jared Rice\altaffilmark{2}, Zhao Zhang\altaffilmark{3}, Bin-Bin Zhang\altaffilmark{3}, and En-Wei
Liang\altaffilmark{1}}

\altaffiltext{1}{Guangxi Key Laboratory for Relativistic Astrophysics, School of Physical Science
and Technology, Guangxi University, Nanning 530004, China; lhj@gxu.edu.edu}
\altaffiltext{2}{Department of Physics, Texas State University, San Marcos, TX 78666, USA}
\altaffiltext{3}{Key Laboratory of Modern Astronomy and Astrophysics (Nanjing University), Ministry
of Education, Nanjing 210093, China}

\begin{abstract}	
Observationally, there is a small fraction of Gamma-ray bursts (GRBs) with prompt emission observed
by Fermi/GBM that are composed of two pulses. Occasionally, the distance to a GRB may be lensed
when a high mass astrophysical object resides in the path between the GRB source and observer. In
this paper, we describe GRB 200716C, which has a two-pulse emission and duration of a few seconds.
We present a Bayesian analysis identifying gravitational lensing in both temporal and spectral
properties, and calculate the time delay ($\Delta t\sim 1.92$ s) and magnification ($\gamma\sim
1.5$) between those two pulses based on the temporal fits. One can roughly estimate the lens mass
to be about $2.4\times 10^{5}~M_{\odot}$ in the rest frame. We also calculate the false alarm
probability for this detection to be about 0.07\% with trial factors, and a present-day number
density of about $808 \rm~Mpc^{-3}$ with an energy density $\Omega\sim 1.4\times 10^{-3}$. If the
first pulse of this GRB near the trigger time is indeed gravitationally echoed by a second pulse,
GRB 200716C may be a short GRB candidate with extended emission.
\end{abstract}


\section{Introduction}
The theory of general relativity (GR) predicts that space is curved by compact objects, and the
phenomenon arising from the deflection of electromagnetic radiation (light rays) toward the mass in
a gravitational field is called gravitational lensing (Blandford \& Narayan 1992). A point mass
gravitational lens magnifies and makes two different images of the source when a massive object is
located close to the line of sight between the observer and a source (see Treu 2010 for a review).
The photons traveling a longer distance will arrive first, but those traversing a shorter path
deeper into the gravitational potential of the lens will have a delayed arrival due to a larger
time dilation. Thus, the gravitationally retarded image is dimmer than the first image (see Section
2 for details). The observational signature of such an effect is an initial pulse followed by a
duplicate echoed pulse. The duration of the time delay between the two emissions depends on the
mass of the gravitational lens and the magnification of the two images (Mao 1992; Paynter et al.
2021). The profile of the light curve of the two images should be similar even with their different
intensities. However, the gravitational lensing process does not change the photon energies, such
that all source images should have the same spectra (Paczynski 1987; Mao 1992).

Gamma-ray bursts (GRBs) are some of the most luminous and active high-energy transients that have
been described since their discovery in 1963 (see Kumar \& Zhang 2015 for a review), and their
spectroscopically measured redshifts span a range from 0.0085 (Galama et al. 1998; Kulkarni et al.
1998) to 9.4 (Cucchiara et al. 2011) with more than $10^4$ observed GRBs. The discovery of
gravitationally lensed optical sources with redshifts ranging from 0.102 to 5.699, implies that
GRBs may be gravitationally lensed occasionally (Paczynski 1986). If this is the case, GRBs play an
important role in searching for evidence of gravitational lensing. Each image with a
gravitationally induced time delay and different magnification can be detected through the observed
burst light curve (Paczynski 1986; Blandford \& Narayan 1992; Kalantari et al. 2021). Based on the
time delay and the ratio of peak flux of the two images, one can roughly estimate the lens mass in
the rest frame (Mao 1992; Paynter et al. 2021; Kalantari et al. 2021).

From an observational point of view, a small fraction of GRBs with prompt emission observed by the
Fermi Gamma-ray Burst Monitor (GBM) are composed of two or more emission episodes with a quiescent
time that may last up to $\sim$100 s in the rest frame (Koshut et al. 1995; Lazzati 2005; Burlon et
al. 2008; Bernardini et al. 2013; Hu et al. 2014; Lan et al. 2018). More interestingly, Lan et al
(2018) performed a systematic analysis of both the spectral and temporal properties of GRBs with
prompt emission observed by Fermi/GBM showing two-episode emission components in the light curves
with quiescent times of up to hundreds of seconds. Statistically speaking they found that the
spectral properties of the two-episode emission components are not significantly different, but
they did not analyze carefully the light curves of those two-episode components. Recently, Paynter
et al. (2021) claimed that they have found a possible signature of a gravitational lens in the
light curve of GRB 950830 with two-episode emission. This could mean that the two-episode emission
signatures are gravitationally lensed images of the same single-episode source. However, they do
not present more details of the spectral properties of the two-episode emission.

One question is whether we can search for robust signatures of gravitational lensing in GRBs that
produce two images within the source-lens-observer geometry and manifest in both the light curves
and spectra. By systematically searching for more than 3000 GRBs observed with both Fermi/GBM and
the Swift Burst Alert Telescope (BAT), we found GRB 200716C with two-episode emission. Its temporal
and spectral properties satisfy the requirements of the theoretical predictions of gravitational
lensing. In this paper, {we show the basic theory of gravitational lensing in \S 2. Then, we
discuss the evidence for gravitational lensing of GRB 200716C based on the observational data. A
comprehensive data reduction and analysis of GRB 200716C is presented in \S 3, and a lens mass
estimate is shown in \S 4. Conclusions are drawn in \S 5 with some additional discussion.

\section{Basic theory of gravitational lensing}
Considering a light ray from a distant source approaching a point mass ($M$), the bend angle
$\alpha$ in the geometric optics limit is given as
\begin{eqnarray}
\alpha=\frac{4GM}{c^{2}b},
\label{bend-angle}
\end{eqnarray}
where $b$ is the impact parameter by denoting the distance of closest approach of the ray to the
mass, and $G$ and $c$ are the gravitational constant and speed of light, respectively. Figure
\ref{fig:cartoon} is a cartoon picture of the point mass gravitational lens geometry. First, let us
label the observer-source distance $D_{\rm os}$, the observer-lens distance $D_{\rm ol}$, and the
lens-source distance $D_{\rm ls}$. By assuming the weak field and thin-lens approximation, one has
$\alpha=\frac{4GM}{c^{2}b}\ll 1$ (i.e., for a small angle) and $b\ll D_{\rm ol}$ which implies
$\theta \ll 1$. Since $\beta < \theta$, $\beta$ is also small angle. Based on the small angle
mathematical geometry of projecting on a vertical line, we can write
\begin{eqnarray}
D_{\rm ls}\alpha+D_{\rm os}\beta=D_{\rm os}\theta
\label{geometry}
\end{eqnarray}
Combining Eq. (\ref{bend-angle}) and Eq. (\ref{geometry}), one can solve the quadratic equation for
$\theta$, and find two solutions,
\begin{eqnarray}
\theta_{\pm}=\frac{1}{2}[\beta \pm (\beta^{2}+\frac{16GM}{c^2}\frac{D_{\rm ls}}{D_{\rm ol} D_{\rm
os}})^{1/2}]
\label{lensed-theta}
\end{eqnarray}
For small angles, one multiplies both sides of Eq. (\ref{lensed-theta}) by $D_{\rm ol}$ to solve
for $b$,
\begin{eqnarray}
b_{\pm}=\frac{1}{2}[\lambda \pm (\lambda^{2}+\frac{16GM}{c^2}\frac{D_{\rm ls}D_{\rm ol}}{D_{\rm os}})^{1/2}].
\label{lensed-image}
\end{eqnarray}
Thus, there will always be two lensed images for a point mass lens (Blandford \& Kochanek 1987).

In order to find out the relationship between time delay ($\Delta t$) and magnification ($\gamma$)
from the unlensed to the lensed system, we define a critical radius (also called the Einstein
radius),
\begin{eqnarray}
r_{\rm cr}=(\frac{4GM}{c^2}\frac{D_{\rm ls}D_{\rm ol}}{D_{\rm os}})^{1/2}
\label{critical-radius}
\end{eqnarray}
inside  which significant magnification takes place because the lensing changes the cross section
but not the surface brightness (Turner et al. 1984). By defining a dimensionless impact parameter
$f=\lambda/r_{\rm cr}$, the Eq.(\ref{lensed-image}) can become
\begin{eqnarray}
b_{\pm}=\frac{r_{\rm cr}}{2}[f\pm \sqrt{f^2+4}].
\label{lensed-image-1}
\end{eqnarray}
The magnification (or the ratio of fluxes of individual images) can be expressed as
\begin{eqnarray}
\gamma=\frac{I_{\rm b_{+}}}{I_{\rm b_{-}}}=\frac{(f^2+2)+f\sqrt{f^2+4}}{(f^2+2)-f\sqrt{f^2+4}}
\label{magnification}
\end{eqnarray}
The time delay is contributed by two effects when the arrival of photons follows the two paths in
Figure \ref{fig:cartoon}. One is geometric due to different path lengths and the other is that two
rays experience different general relativistic time dilations when the two paths traverse different
gravitational potentials (Weinberg 1972). Thus, the time delay can be given as
\begin{eqnarray}
\Delta t=\frac{D_{\rm ol}D_{\rm ls}}{2D_{\rm os}}(\alpha^2_{-}-\alpha^2_{+})+\frac{2GM_{\rm z}}{c^3}
\rm ln(\frac{b^2_{+}}{b^2_{-}})
\label{time-delay-1}
\end{eqnarray}
By invoking Eq. (\ref{magnification}), one can rewrite the time delay as (Krauss \& Small 1991)
\begin{eqnarray}
\Delta t=\frac{2GM_{\rm z}}{c^3}[\frac{\gamma-1}{\sqrt{\gamma}}+\rm ln(\gamma)]
\label{time-delay-2}
\end{eqnarray}
where $M_{\rm z}=M(1+z)$ is the redshifted lens mass.

GRBs have a good temporal resolution in the $\gamma$-ray band, and the time delay and magnification
between the two images can be observed by considering both the difference in geometric path and the
relative difference in gravitational field strength. Thus it is easy to estimate the mass of the
gravitational lens:
\begin{eqnarray}
M_{z}=\frac{c^3\Delta t}{2G}(\frac{\gamma-1}{\gamma}+\rm ln(\gamma))^{-1}.
\label{masslens}
\end{eqnarray}

\section{Data reduction and analysis}
In order to test how many of the GRBs observed by Fermi/GBM are potentially gravitationally lensed,
as of 2021 July, we downloaded the original GBM data (12 NaI and 2 BGO detectors) of 3035 GRBs from
the public science support center at the official Fermi
website\footnote{http://fermi.gsfc.nasa.gov/ssc/data/}. We employ the Bayesian Block algorithm to
identify the light curves, and extract the spectrum using our automatic code ``{\em McSpecfit}''.
Please refer to our previous paper (Lan et al. 2018) for more details on data analysis with the
Bayesian Block algorithm, and to Zhang et al. (2018) for details on the spectral fitting. There are
two criteria adopted for our sample selection. First, the GRB prompt emission must have two-episode
(or more) emission, and the signal-to-noise ratio (S/N) of the emission episodes should be greater
than 3$\sigma$. Second, the spectra of the two-episode (or more) emission should be similar to each
other. After searching 3035 GRBs, we find that only GRB 200716C satisfies our criteria.

\subsection{The basic observations of GRB 200716C}
GRB 200716C triggered Swift/BAT, Insight-HXMT, and Fermi/GBM. Due to the lack of public
Insight-HXMT data, in this section we focus on introducing the prompt emission of GRB 200716C
observed by Swift/BAT and Fermi/GBM, as well as the afterglow (both X-ray and optical)
observations.

GRB 200716C triggered the BAT at 22:57:41 UT on 16 July 2020 (Ukwatta et al. 2020). We downloaded
the BAT data from the Swift website\footnote{$\rm
https://www.swift.ac.uk/archive/selectseq.php?tid=00982707\&source=obs$}, and use the standard
HEASOFT tools (version 6.28) to process the BAT data. For more details of the analysis, please
refer to Sakamoto et al. (2008); Zhang et al. (2009); and L\"{u} et al. (2020). The light curves in
different energy bands are extracted with the time-bin size 8 ms. Then, we calculate the cumulative
distribution of the source counts using the arrival time. The light curve shows two prominent peaks
with a duration of about $5.3$ s in 15-150 keV (see Figure \ref{fig:BATLC}), but weak activity is
still visible until about 90 seconds.

At 22:57:41.18 UT on 2020 July 16, the GBM was triggered and located GRB 20716C (Veres et al.
2020). GBM has 12 sodium iodide (NaI) and two bismuth germanate (BGO) scintillation detectors
covering the energy range from 8 keV to 40 MeV (Meegan et al. 2009). We downloaded the
corresponding Time-Tagged-Event data from the public data site of Fermi/GBM\footnote{$\rm
https://heasarc.gsfc.nasa.gov/FTP/fermi/data/gbm/triggers/$}. For more details of the light-curve
data reduction procedure refer to Zhang et al. (2016). The light curves of the n0 and b0 detectors
with 8 ms and 64 ms time bins are shown in Figure \ref{fig:GBMLC}, and consist of two pulses with a
duration $3.3$ s in 50-300 keV. There is no significant weak emission after the second pulse in the
GBM temporal analysis.

The X-ray telescope (XRT) began observing the field at 22:59:04.2 UT, 82.9 s after the BAT trigger
(Ukwatta et al. 2020). We made use of the public data from the Swift archive \footnote{$\rm
https://www.swift.ac.uk/xrt\_curves/00982707$}(Evans et al. 2009). The X-ray light curve seems to
be a power-law decay until $\sim 10^{5}$ s with decay slope $\alpha_0=1.55\pm 0.02$ (see Figure
\ref{fig:Xray}). Kann et al. (2020) observed the position of the afterglow with the 1.23 m Calar
Alto telescope starting with the second Swift orbit and found that the decays follow a broken power
law with decay slopes $\alpha_1=0.8\pm 0.04$, $\alpha_2=5.5\pm 1.3$, and break time $t_{\rm
b}=(3.8\pm0.26)\times 10^{4}$ s (see Figure \ref{fig:Xray}).

\subsection{Light-curve fits of GRB 200716C}
The light curve of GRB prompt emission with pulses is usually described with the fast-rise
exponential-decay (FRED) model (Norris et al. 1996). In order to test the consistency of structure
for the two pulses, we also employ the FRED model to fit the pulses of GRB 200716C. By invoking the
public code from Paynter et al. (2021), we used the same method from Paynter et al. (2021) to fit
the light curve\footnote{For more details of this method and public code, please refer to Paynter
et al. (2021).}. Here, we adopt two approaches to fit the data. Firstly, we used the same
parameters (except the peak time and normalization) of one FRED model to fit the two pulses and
obtain the values ln($\mathcal{Z}_{\rm L}$), if we believe they are gravitationally lensed (called
``FL''). Next, we used two FRED models to fit the two pulses with different parameters to get the
values ln($\mathcal{Z}_{\rm NL}$), if they are independent of one another (called ``FF''). The
light curve of a statistically significant gravitational lensing candidate GRB 200716C is shown in
Figure \ref{fig:BATLC}. The reconstructed curves of the best model fit are plotted in black. We
also present the difference between the true light curve and the posterior predictive curve in
different energy channels. We find that the residuals are consistent with zero, which means the
lens model we selected is a good one. On the other hand, in order to determine which model is
preferred by the data we also calculate the Bayesian evidence for each model with the Bayes factor
(ln$BF$), which is defined as ln$(BF)= (\rm ln(\mathcal{Z}_{\rm L})- \rm ln(\mathcal{Z}_{\rm
NL}))$. A ln$(BF)$ that is larger than 8 is considered strong evidence for supporting one model
over another (Thrane \& Talbot 2019; Paynter et al. 2021).

We separate the Swift/BAT light curves into four available broadband energy channels, and
independently calculate the value of ln$(BF)$ in those four channels (see Table 1). We find that
the values of ln$(BF)$ are between $-0.1$ and $7.0$ in each channel, and the total ln$(BF)$ value
from each of the channels is about $15.24$ in favour of the lensing hypothesis. This is strong
statistical evidence supporting the lensing hypothesis.

Similar to the pulse fitting of Swift/BAT data, we also apply the FRED model to fit the the
Fermi/GBM data. The reconstructed curves of the best model fits are plotted in black (see Figure
\ref{fig:GBMLC}). The residual in the different energy channels are also consistent with zero,
indicating that the lens model is the best one. Here, we calculate the Bayes factor in four
available energy channels (see Section 4) with 8 ms and 64 ms time bins, respectively. For the 8 ms
time bin, the values of ln$(BF)$ are between $0.5$ and $9.0$ in each channel (see Table 1), and the
total ln$(BF)$ value from each of the channels is about $19.94$ in favor of the lensing hypothesis.
But for the 64 ms time bin, the ln$(BF)$ is $-0.5$ during the first energy channel and in the other
three channels it ranges from $4.0$ to $9.0$. The total ln$(BF)$ value from each of the channels is
about $19.56$, which is close to the value of ln$(BF)$ for the 8 ms time bin. This suggests that
the total ln$(BF)$ value for each energy channel seems to be not dependent on the time resolution.
At the least this is also strong statistical evidence supporting the lensing hypothesis.

\subsection{Extracting and fitting the spectrum of GRB 200716C}
We do not extract the spectrum of GRB 200716C observed by BAT due to its narrow energy band, but
focus on the wide energy band in GBM. We extract the time-averaged spectrum of the first (time
interval $(-0.3-1.9)$ s) and second (time interval $(1.9-4.1)$ s) pulses of GRB 200716C,
respectively. The background spectra are extracted from the time intervals before and after those
two pulses and modeled with an empirical function (Zhang et al. 2011). The spectral fitting is
performed by using a Markov Chain Monte Carlo (MCMC) method with our automatic code ``{\em
McSpecfit}'' in Zhang et al. (2018). We adopted several spectral models, which we usually select to
test the spectral fitting of a burst, i.e., power law (PL), cutoff power law (CPL), Band function
(Band), and Blackbody (BB), as well as combinations of any two models. Then, we compare the
goodness of the fits of the two pulses, respectively (see Table 2). Invoking the Bayesian
information criteria (BIC; L\"{u} et al. 2017), we find that the CPL model is the best model that
adequately describes the observed data. The CPL model fit is shown in Figure \ref{fig:SpecGBM}, as
well as the parameter constraints of the fit. For the first pulse, it gives peak energy $E_{\rm
p,1}=(524\pm 97)$ keV, and a lower energy spectral index of $\alpha_{1}=0.96\pm0.05$. For the
second pulse, one has $E_{\rm p,2}=(566\pm 164)$ keV, and $\alpha_{2}=0.98\pm0.08$. The best-fit
parameters of the CPL fits and other models are listed in Table 2.

Within the error range in the spectral data, the spectral properties of the two pulses in the CPL
model are consistent with one another. This consistency is a the prediction of the lensing
hypothesis. Based on the above analysis, both the light curve and spectral properties support that
GRB 200716C is gravitationally lensed.

\section{Estimating the lens mass of GRB 200716C}
In order to determine whether the two pulses of GRB 200716C are a false alarm, based on the method
of Paynter et al. (2021), we also calculate the false alarm probability $1-p_{\rm
lens}=\frac{1}{1+\rm ln(BF)/N}$, where $N=3035$ is total number of GRBs observed by Fermi/GBM. One
has $1-p_{\rm lens}=7.3\times 10^{-4}$ with the 8 ms time bin. In other words, the false alarm
probability for this detection is about 0.07\% with trial factors. Moreover, we also calculate that
number density is about $808\rm~Mpc^{-3}$ with an energy density $\Omega\sim 1.4\times 10^{-3}$ by
assuming a redshift for GRB 200716C of $z=0.348$ (D'Avanzo \& CIBO Collaboration. 2020) and the
average redshift of GRBs observed by Swift $z\sim 2.2$ (Xiao \& Schaefer 2011).

The gravitational lens will not change the  photon energies when the photons travel close to
compact objects, which means that all wavelengths of the light curve are equally affected by
gravitational fields. In other words the time delay of different pulses is independent of the
photon energy and it should be the same in different energy channels. Also, the gravitational
magnification of each image is identical for every wavelength. In order to test this hypothesis
with the observed data, we separate the Swift/BAT and Fermi/GBM light curves into four available
broadband energy channels, respectively\footnote{The light curve of Swift/BAT is divided into four
energy channels: 15-25 keV, 25-50 keV, 50-100 keV, and 100-350 keV. The Fermi/GBM light curve is
separated into 8-44 keV, 44-100 keV, 100-250 keV, and 250-900 keV.}.

Based on the light-curve fits for each energy channel and adopting a method similar to Paynter et
al. (2021), one can easily to calculate the time delay and magnification. For Swift/BAT data, we
roughly calculate $\Delta t \sim 1.93$ s and $\gamma \sim 1.54$. For the 8 ms time bin of Fermi/GBM
data, one has $\Delta t \sim 1.92$ s and $\gamma \sim 1.49$. For the 64 ms time resolution, one has
$\Delta t\sim 1.92$ s and $\gamma \sim 1.52$. This indicates that both time delay and magnification
are also independent of the time resolution. Figure \ref{fig:ratio} shows the peak flux ratio as a
function of energy channels for prompt emission observed by Swift/BAT and Fermi/GBM (8 ms and 64 ms
time bin). The ratio seems to be consistent across the different energy channels and time bins. By
invoking the Eq. (\ref{masslens}), as well as adopting $\Delta t \sim 1.92$ s and $\gamma \sim
1.5$, one can roughly estimate that the lens mass in the rest frame is about $2.4\times
10^{5}~M_{\odot}$. There are several astrophysical objects within this mass range, such as globular
clusters, diffuse galaxies, dark matter, and black holes (Paynter et al. 2021).

If GRB 200716C was lensed by a globular cluster, the estimated cosmic energy density of globular
clusters $\Omega_{\rm gc}\sim 8\times 10^{-6}$ should be consistent with that of predictions.
However, it is inconsistent as we infer energy densities much larger than that of globular clusters
(see Figure \ref{fig:Number-density}). If the astrophysical object is a diffuse galaxy then it
should have strong $\gamma$-ray and radio emission, which is inconsistent with current observations
(Mihos et al. 2005). The other possible astrophysical object is an intermediate-mass black hole
(Paynter et al. 2021), but whether black holes in this mass range exist remains an open question.
By comparison with the result of Paynter et al. (2021), we find that the inferred lens mass of GRB
200716C is about 4 times higher than that of GRB 950803, and the inferred energy density of GRB
200716C is also about 3 times larger than that of GRB 950803. This result is consistent with that
of Paynter et al. (2021).

\section{Conclusion and discussion}
GRB 200716C was observed by Swift, Fermi, and Insight-HXMT to have a duration of few seconds. The
prompt emission of this GRB consists of two pulses and weak emission  (called ``extended emission")
lasting $\sim 90$ s after the second pulse is visible in the Swift/BAT, but not visible in the
Fermi/GBM temporal analysis. In this paper, we presented a comprehensive analysis of its temporal
and spectral data, and tested whether the first pulse of GRB 200716C near the trigger time is
indeed gravitationally echoed by a second pulse, indicating that both pulses are gravitationally
lensed images of the same single source pulse.

Firstly, we separated the Swift/BAT and Fermi/GBM light curves into four available broadband energy
channels, respectively. The FRED model is invoked to fit the profile of two pulses in each channel
by adopting the public code from Paynter et al. (2021). In Figure \ref{fig:BATLC}, we plot the
light curve of 200716C observed by Swift/BAT, as well as the FRED model fits in different energy
channels. The model fit is used to subtract from the true observed light curve and obtain the
residuals. We find that the residuals are consistent with zero, which means that the lens model we
selected is favored. Then, we independently calculate the Bayesian evidence for each model with
Bayes factor (ln$(BF)$). We find that the total ln$(BF)$ value from each of the channels is about
19 for BAT and GBM (even with different time resolution). This value is much larger than 8, and so
the lensing hypothesis is favored. It is also independent of the time resolution of the prompt
emission. Moreover, we also extract the spectral parameters by using the MCMC method with our
automatic code ``{\em McSpecfit}'' in Zhang et al. (2018). Several spectral models (PL, CPL, Band,
and BB), or even combinations of any two models, are selected to fit. We find that the CPL model is
the best one that adequately describes the observed data by comparing the goodness of the fits of
the two pulses, respectively. Both the $E_{\rm p}$ and $\alpha$ values of those two pulses are
consistent with one another within the error range. This consistency is a prediction of the lensing
hypothesis and is strong statistical evidence to support for the lensing hypothesis of GRB 200716C.

One basic question is whether the lensing signal from GRB 200716C is a false alarm. In order to
test this question, we calculate the false alarm probability for this detection, which is about
0.07\% with trial factors based on the method of Paynter et al. (2021). By adopting the redshift of
GRB 200716C to be $z=0.348$ and the average redshift of GRBs observed by Swift to be $z\sim 2.2$,
we estimated the number density as $808\rm~Mpc^{-3}$ with an energy density $\Omega\sim 1.4\times
10^{-3}$. On the other hand, we adopted a method similar to Paynter et al. (2021) and after making
light-curve fits for each energy channel, we calculated the time delay and magnification of the
pulses to be $\Delta t \sim 1.92$ s and $\gamma \sim 1.5$, respectively. We find that the time
delay and magnification of the two pulses are independent of the time resolution of the light
curve. The inferred lens mass is about $2.4\times 10^{5}~M_{\odot}$, which is a mass consistent
with several astrophysical objects such as globular clusters, diffuse galaxies, dark matter, and
black holes (Paynter et al. 2021). However, the globular clusters and diffuse galaxies seem
unlikely to be the candidate astrophysical objects. The black hole is a potential candidate, but
more observations are needed to confirm this in the future.

Upon finishing this paper, our attention was drawn to Wang et al. (2021), who performed an
independent analysis on GRB 200716C to discuss the same points. We find that there are two points
of difference between this paper and Wang et al. (2021). First, the spectral fitting results of the
two pulses are different, which may be caused by the different time interval selected and different
fitting methods for the two papers. Several spectral models (PL, CPL, Band, and BB), or even
combinations of any two models, are selected as fitting functions in our paper by using the MCMC
method in our automatic code ``{\em McSpecfit}''.  Wang et al. (2021) used only the Band function
and CPL models to do fits but did not invoke an MCMC method to do that. Second, the estimated lens
mass is slightly different for the two papers, but within the same order. The reason for this may
be the selection of different time delay and magnification values. We used the average time delay
and magnification values of Fermi and Swift in different energy bands to roughly estimate the lens
mass, but Wang et al. (2021) presented the time delay and magnification values in each energy band
and then estimated the lens mass.

If the GRB 200716C is indeed gravitationally lensed, the total duration of the prompt emission of
this GRB should be the duration of any one pulse. If this is the case, then GRB 200716C should be a
typical short-duration GRB with extended emission. Wang et al. (2021) claim that the $E_{\rm
p}-E_{\rm \gamma,iso}$ of GRB 200716C is located in the population of typical short GRBs, even for
individual pulses by assuming a possible\footnote{D'Avanzo \& CIBO Collaboration (2020) reported
the presence of an extended object classified as a galaxy at a position consistent (within $\sim
1''$) with the one reported for the optical afterglow of GRB 200716C. This galaxy is the possible
host galaxy of GRB 200716C.} redshift $z=0.348$ (D'Avanzo \& CIBO Collaboration 2020). At least for
this case, due to the lack of accurate information on emission or absorption lines in the spectrum,
we only can only find some indirect evidence for the gravitational lensing of GRB 200716C. The
``Smoking gun" of gravitational lensing of GRBs is not only the consistency of the temporal and
spectral properties with predictions from gravitational lensing, but the consistency with some
empirical relations, and indeed accurate information of its host galaxy with two images. With the
improvement of detection technology, we encourage observers in the future to invoke large optical
telescopes to follow-up, especially for the GRBs with two-pulse emission. Moreover the light-curve
behaviors between the X-ray and optical are quite different, so it makes this an event of interest.
We also need to carry out a follow-up in the future.

Since the lensing signal could be due to similar-looking pulses of the GRB, the lensing hypothesis
is one possible explanation for the double-pulse structure of GRB 200716C. On the other hand, the
double pulse associated with a GRB 200716C-like event or even repeating pulses could be an
intrinsic feature of the GRB prompt emission (Veres et al. 2021). In this case, it would be
impossible to confidently detect lensing by looking at the similarity of the pulses.

\begin{acknowledgements}
We acknowledge the use of the public data from the Swift data and Fermi data archive. This work is
supported by the National Natural Science Foundation of China (grant Nos. 11922301, and 12133003),
the Guangxi Science Foundation (grant Nos. 2017GXNSFFA198008, and AD17129006), the Program of Bagui
Young Scholars Program (LHJ), and special funding for Guangxi distinguished professors (Bagui
Yingcai and Bagui Xuezhe).
\end{acknowledgements}


\begin{table}
\label{tab:Table1} \centering \caption{The Bayes factor of the fits in different energy bands
observed by Swift/BAT and Fermi/GBM. In the model, FL=``lens'', and FF=``no lens''.}

\tablewidth{100pt} \tabletypesize{\footnotesize}
\begin{tabular}{c|c|c|c|c}
\hline
Instrument              & Energy Channels                & ln(BF)                & Model & ln($\mathcal{Z}$)
\\ \hline
\multirow{8}{*}{Swift/BAT} & \multirow{2}{*}{(15-25) keV}    & \multirow{2}{*}{-0.13}  & FL    &
-77.49$\pm$0.32     \\
\cline{4-5}
                       &                                &                       & FF    & -77.36$\pm$0.33
                       \\
                       \cline{2-5}
                       & \multirow{2}{*}{(25-50) keV}    & \multirow{2}{*}{2.56} & FL    & -100.01$\pm$0.40
                       \\
                       \cline{4-5}
                       &                                &                       & FF    & -102.57$\pm$0.44
                       \\
                       \cline{2-5}
                       & \multirow{2}{*}{(50-100) keV}   & \multirow{2}{*}{5.83} & FL    & -86.50$\pm$0.40
                       \\
                       \cline{4-5}
                       &                                &                       & FF    & -92.33$\pm$0.49
                       \\
                       \cline{2-5}
                       & \multirow{2}{*}{(100-350) keV}  & \multirow{2}{*}{6.98} & FL    & -54.20$\pm$0.38
                       \\
                       \cline{4-5}
                       &                                &                       & FF    & -61.18$\pm$0.42
                       \\ \hline
\multirow{8}{*}{Fermi/GBM (8 ms)} & \multirow{2}{*}{(8-44) keV}     & \multirow{2}{*}{0.59}    & FL    &
-1326.21$\pm$0.22     \\ \cline{4-5}
                       &                                &                       & FF    & -1326.80$\pm$0.25
                       \\ \cline{2-5}
                       & \multirow{2}{*}{(44-100) keV}   & \multirow{2}{*}{7.76}    & FL    &
                       -1121.57$\pm$0.23     \\ \cline{4-5}
                       &                                &                       & FF    & -1129.33$\pm$0.26
                       \\ \cline{2-5}
                       & \multirow{2}{*}{(100-250) keV}  & \multirow{2}{*}{2.7}    & FL    &
                       -1158.81.21$\pm$0.24     \\ \cline{4-5}
                       &                                &                       & FF    & -1161.51$\pm$0.27
                       \\ \cline{2-5}
                       & \multirow{2}{*}{(250-1000) keV} & \multirow{2}{*}{8.89}   & FL    &
                       -872.57$\pm$0.21
                       \\ \cline{4-5}
                       &                                &                       & FF    & -881.46$\pm$0.25
                       \\ \hline
\multirow{8}{*}{Fermi/GBM (64 ms)} & \multirow{2}{*}{(8-44) keV}     & \multirow{2}{*}{-0.5}    & FL    &
-273.13$\pm$0.22     \\ \cline{4-5}
                       &                                &                       & FF    & -272.63$\pm$0.24
                       \\ \cline{2-5}
                       & \multirow{2}{*}{(44-100) keV}   & \multirow{2}{*}{4.76}    & FL    &
                       -261.45$\pm$0.23     \\ \cline{4-5}
                       &                                &                       & FF    & -266.21$\pm$0.26
                       \\ \cline{2-5}
                       & \multirow{2}{*}{(100-250) keV}  & \multirow{2}{*}{8.64}    & FL    &
                       -272.10$\pm$0.23     \\ \cline{4-5}
                       &                                &                       & FF    & -280.74$\pm$0.27
                       \\ \cline{2-5}
                       & \multirow{2}{*}{(250-1000) keV} & \multirow{2}{*}{6.66}   & FL    &
                       -216.33$\pm$0.21
                       \\ \cline{4-5}
                       &                                &                       & FF    & -222.99$\pm$0.24
                       \\ \hline
\hline
\end{tabular}
\end{table}


\begin{table}
\label{tab:Table2} \centering \caption{Spectral fitting results of GRB 200716C with different
models} \tablewidth{100pt} \tabletypesize{\footnotesize} \large \resizebox{\textwidth}{20mm}{
\begin{tabular}{cc|cccccc|cccccc}
\hline
&Model & & & Pulse-1  &  & & & &  &Pulse-2 &  & &    \\ \hline
&   &$\Gamma$ &$\alpha$  &$\beta$ &$E_{p,1}$ &kT &BIC   &$\Gamma$ &$\alpha$  &$\beta$ &$E_{p,2}$ &kT &BIC   \\ \hline
&BB	 	  & 	& 	& 	& 	& $50\pm2$	&774		& 	& 	& 	& 	& $51.87\pm2.22$	&728\\ \hline
&CPL	  &	& $0.96\pm0.05$ 	& 	& $ 523\pm 97$	& 	&342	 	& 	& $0.98\pm0.08$	& 	& $ 566\pm 163$	& 	 &529\\ \hline
&CPL+BB	  &	&  $1.02\pm0.16$	& 	& $ 306\pm 98$	& $ 128\pm 2$	&349		& 	 &$0.56\pm0.29$ 	& 	& $ 320\pm 122$	& $ 9.13\pm 1.43$	&536\\ \hline
&CPL+PL	  & $2.11\pm3.39$	& $0.88\pm0.34$	& 	& $ 456\pm 209$	& 	&353		& $8.6415\pm24.23$	 &$0.98\pm0.47$ 	& 	& $ 576\pm 350$	& 	&540\\ \hline
&Band	  & 	& $-0.96\pm0.05$	& $ -9.3\pm 3804$	& $ 522\pm 97$	& 	&348		& 	& $-0.97\pm0.08$	 & $ -8.5\pm 5671$	& $ 567\pm 174$	& 	&535\\ \hline
&Band+BB  & 	& $-0.95\pm0.06$	& $ -9.3\pm 3802$	& $ 518\pm 101$	& $ 0.84\pm 2.42$	&359		 & 	& $-0.57\pm0.29$	& $ -6.8\pm 617.$	& $ 325\pm 125$	&
$
9.14\pm 1.43$	&542\\ \hline
&Band+PL  & $ 9.34\pm 28.8$	& $-0.95\pm0.061$	& $ -9.3\pm 3795$	& $ 520\pm 102$	& 	 &359		& $ 2.43\pm 5.09$	& $-0.97\pm0.077$	& $ -8.6\pm 4907$	& $
562\pm 184$	& 	&546\\ \hline

\end{tabular}
}
\end{table}


\begin{figure}
\centering
\includegraphics [angle=0,scale=0.7] {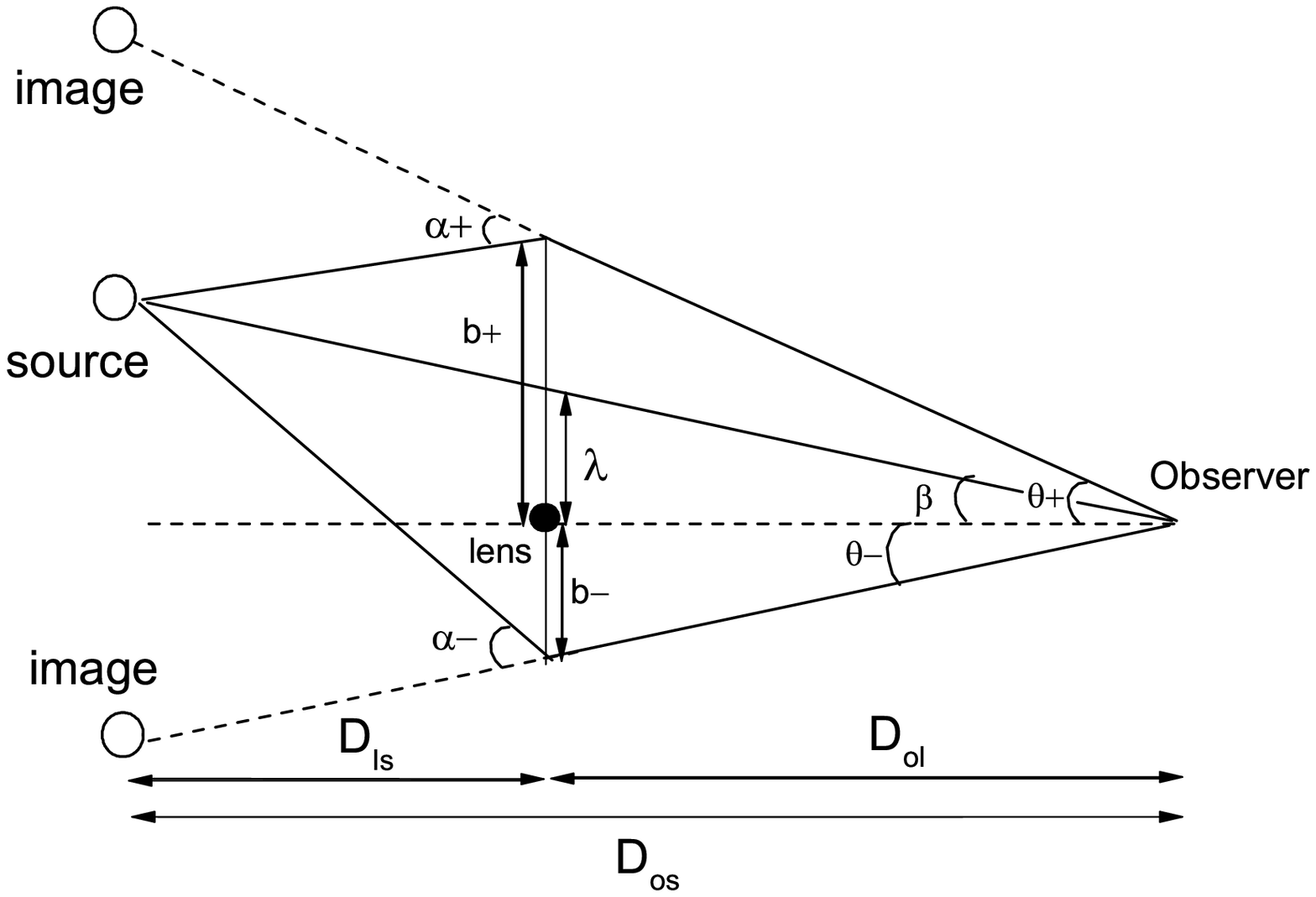}
\caption{Cartoon picture of the geometry of gravitational lensing.}
\label{fig:cartoon}
\end{figure}
\begin{figure}
\centering
\includegraphics [angle=0,scale=0.8] {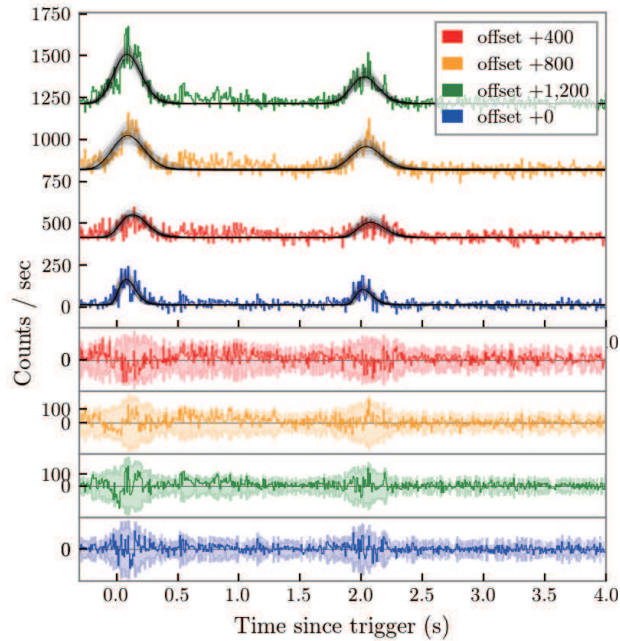}
\caption{Light curve of prompt emission for gravitational lensing GRB 200716C observed by Swift/BAT in the top four panels.
Different colors indicate different energy channels: red, (15-25) keV; yellow, (25-50) keV; green,
(50-100) keV; and blue, (100-350) keV. The solid black lines are the best fit with the empirical function (FRED).
The bottom four panels correspond to residuals that show the data after the template has been subtracted
for different energy channels. The colored shaded regions are the 1$\sigma$ standard statistical error. These panels seem to show
that the lens model is a reasonable fit.}
\label{fig:BATLC}
\end{figure}

\begin{figure}
\centering
\includegraphics [angle=0,scale=0.8] {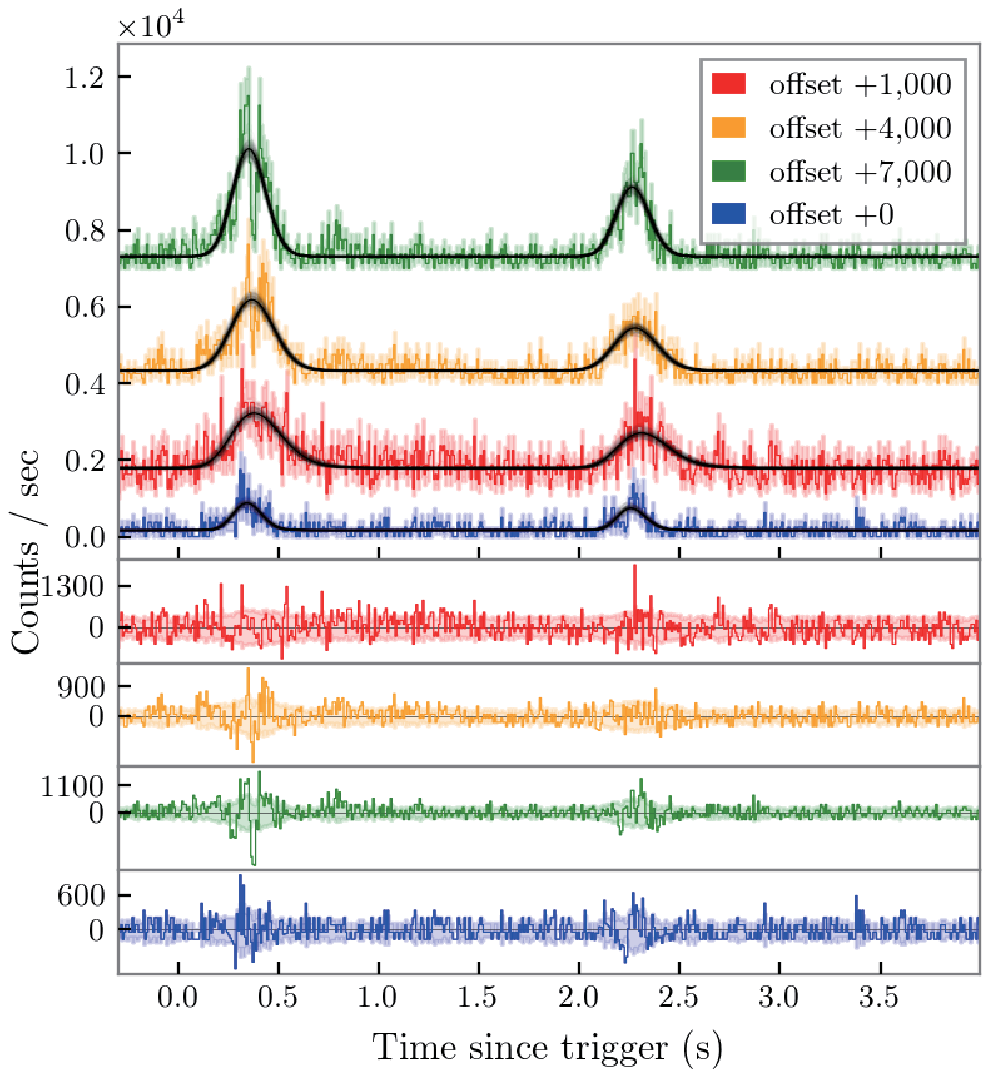}
\includegraphics [angle=0,scale=0.8] {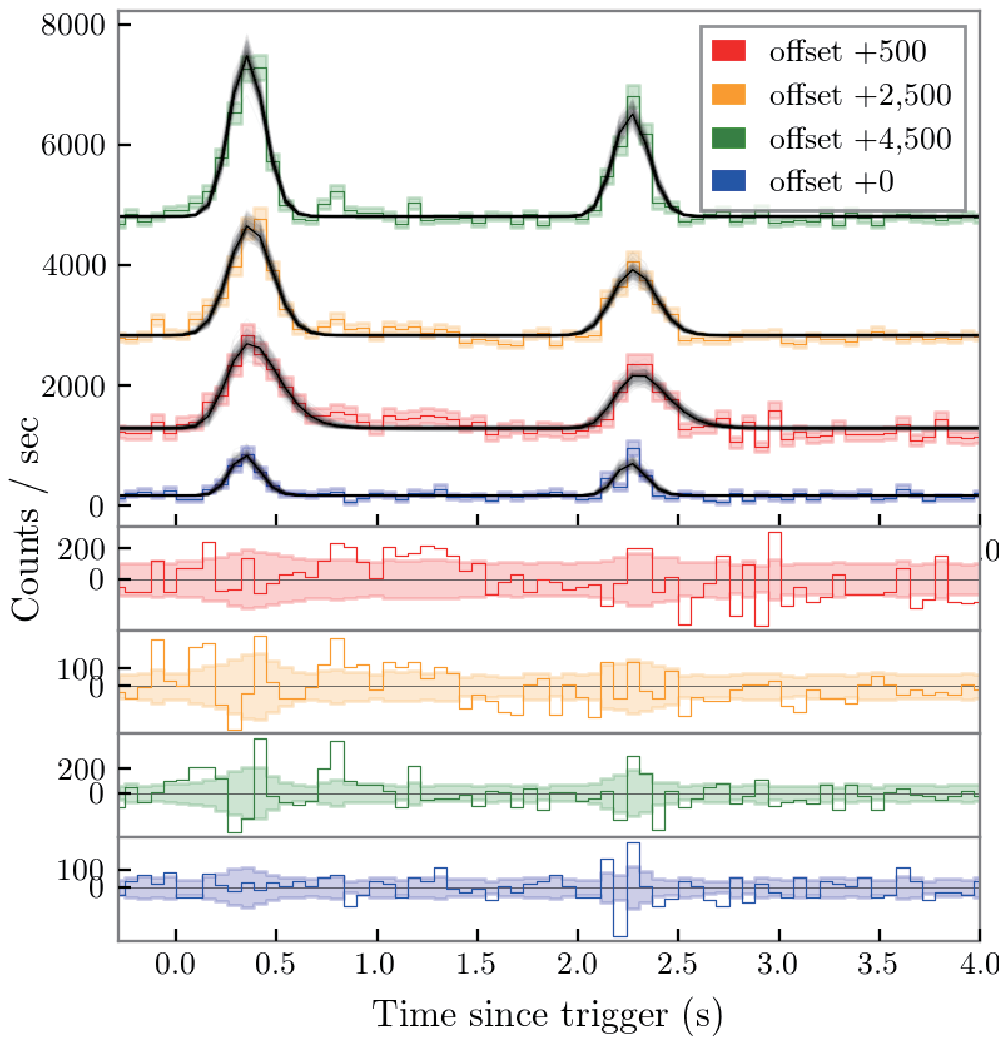}
\caption{Similar to Figure \ref{fig:BATLC}, but observed by Fermi/GBM and different
energy channels: red, (8-44) keV; yellow, (44-100) keV; green,
(100-250) keV; and blue, (250-900) keV. The left and right panels are the 8 ms and 64 ms time bins, respectively.
The residuals of different the energy channels are consistent with zero, indicating that
the lens model is a good one for the data.}
\label{fig:GBMLC}
\end{figure}
\begin{figure}
\centering
\includegraphics [angle=0,scale=0.8] {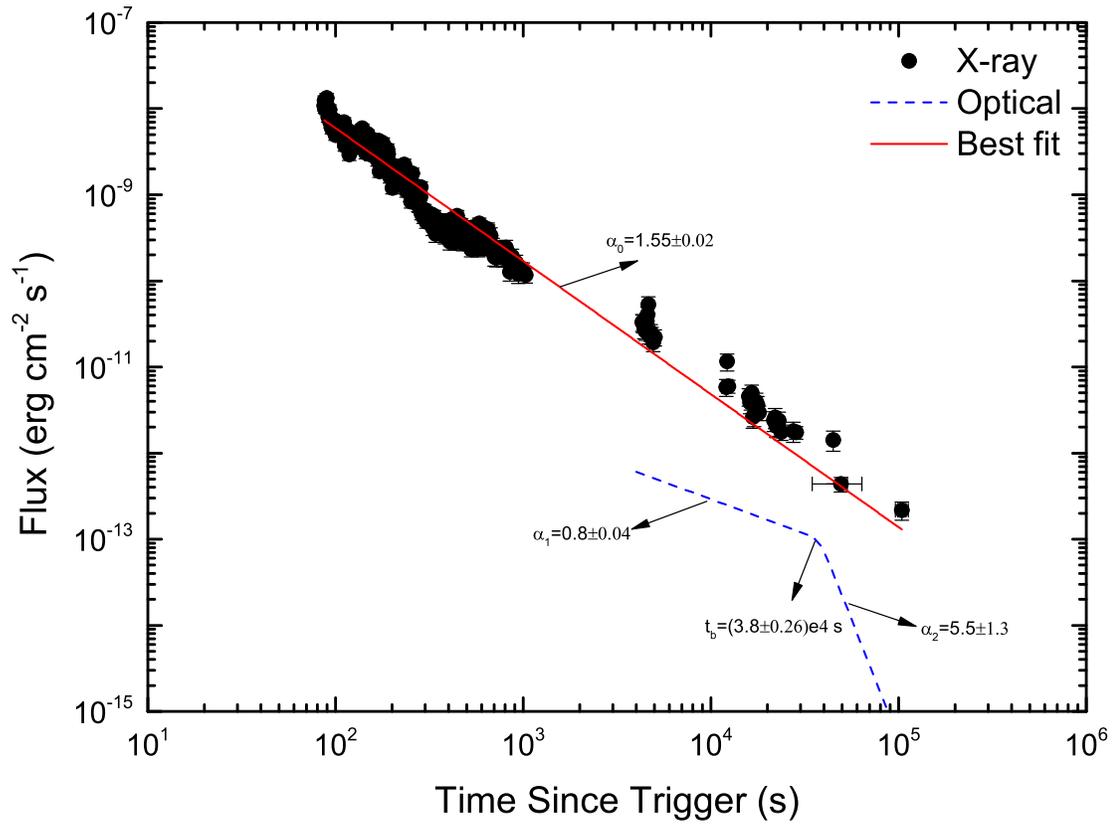}
\caption{X-ray and optical light curves of GRB 200716C.
The solid line is the best fit with the power-law model.}
\label{fig:Xray}
\end{figure}

\begin{figure}
\centering
\includegraphics [angle=-90,width=0.4\textwidth] {f5a.eps}
\includegraphics [angle=-90,width=0.4\textwidth] {f5c.eps}
\includegraphics [angle=0,width=0.4\textwidth] {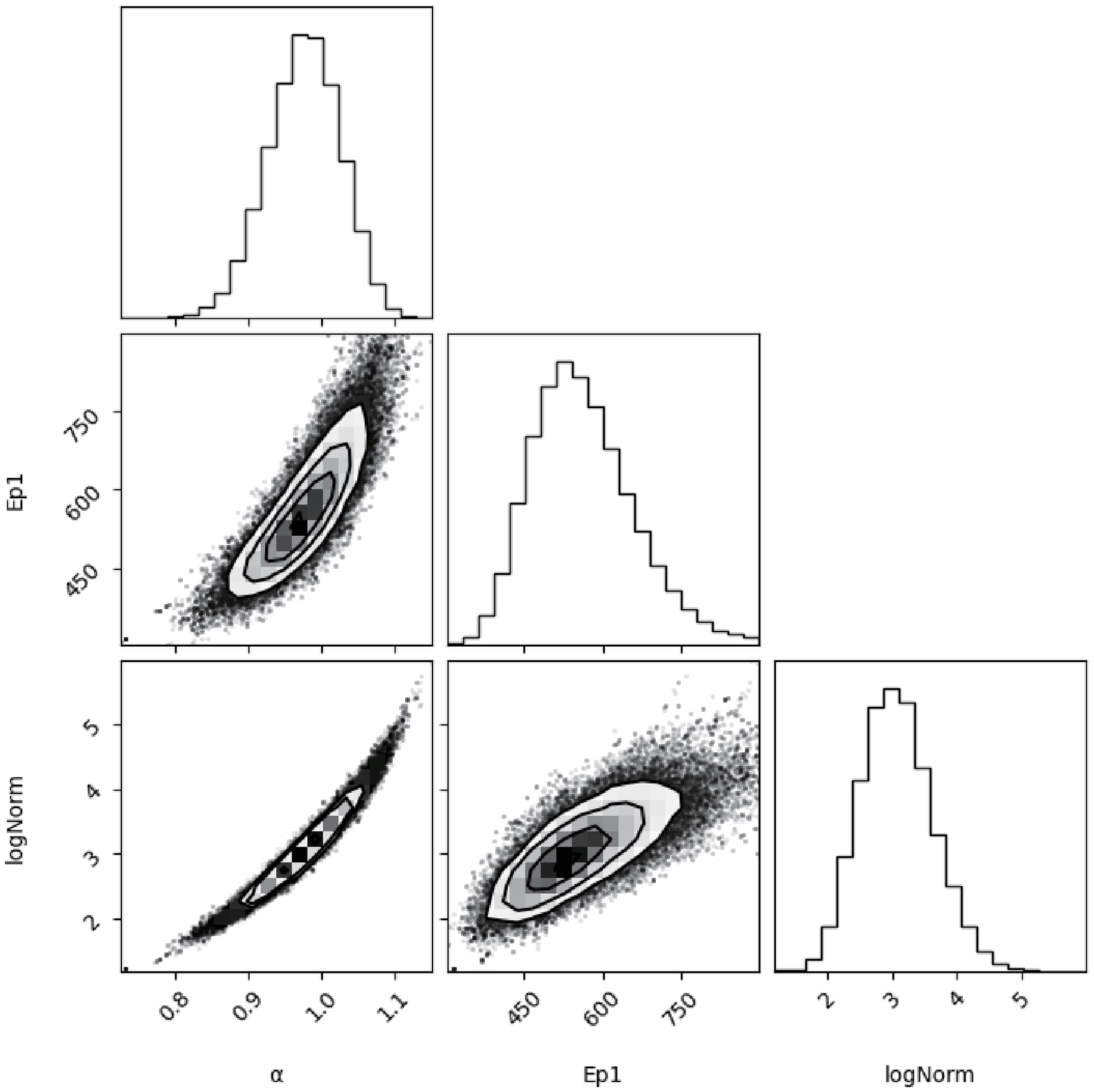}
\includegraphics [angle=0,width=0.4\textwidth] {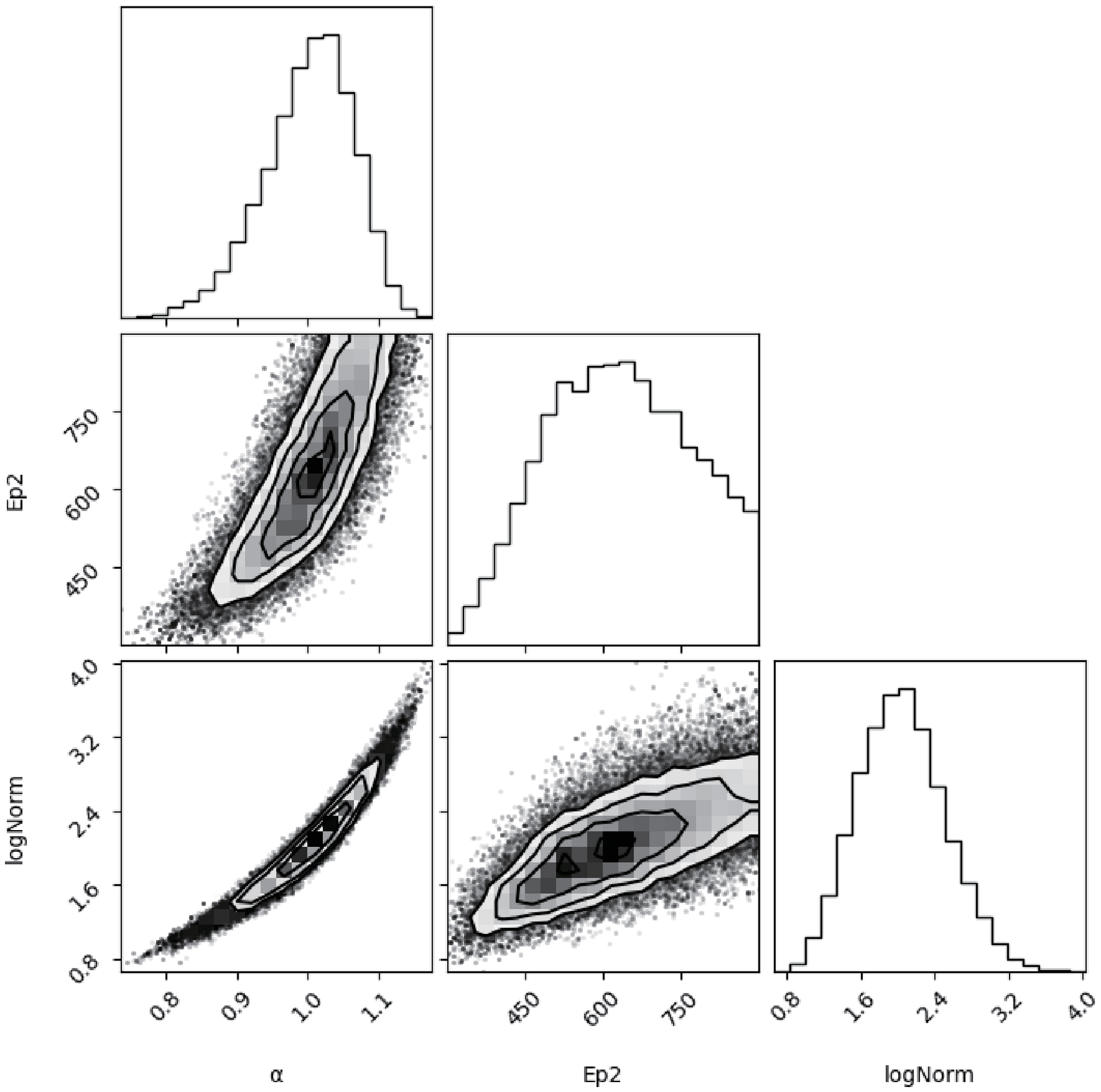}
\caption{Spectral fits of GRB 200716C with the cutoff power-law model for Fermi/GBM.
The $\nu F_{\nu}$ spectrum and parameter constraints of the CPL fit for the first
(left panels) and second pulses (right panels), respectively. Histograms and contours in the
corner plots show the likelihood map of constrained parameters by using
our McSpecFit package. The solid black circles are the 1$\sigma$, 2$\sigma$, and 3$\sigma$
uncertainties, respectively.}
\label{fig:SpecGBM}
\end{figure}
\begin{figure}
\centering
\includegraphics [angle=0,scale=0.8] {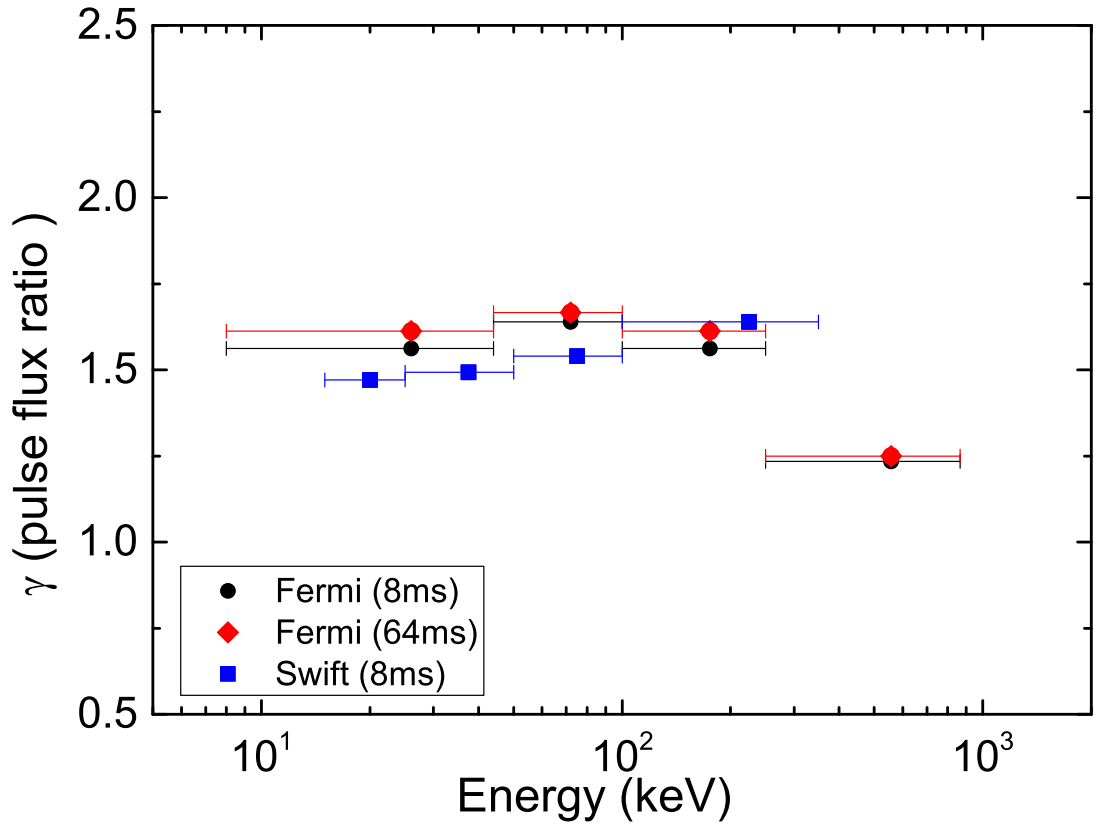}
\caption{Peak flux ratio between the two pulses as a function of energy channels for GRB 200716C.}
\label{fig:ratio}
\end{figure}

\begin{figure}
\centering
\includegraphics [angle=0,scale=0.1] {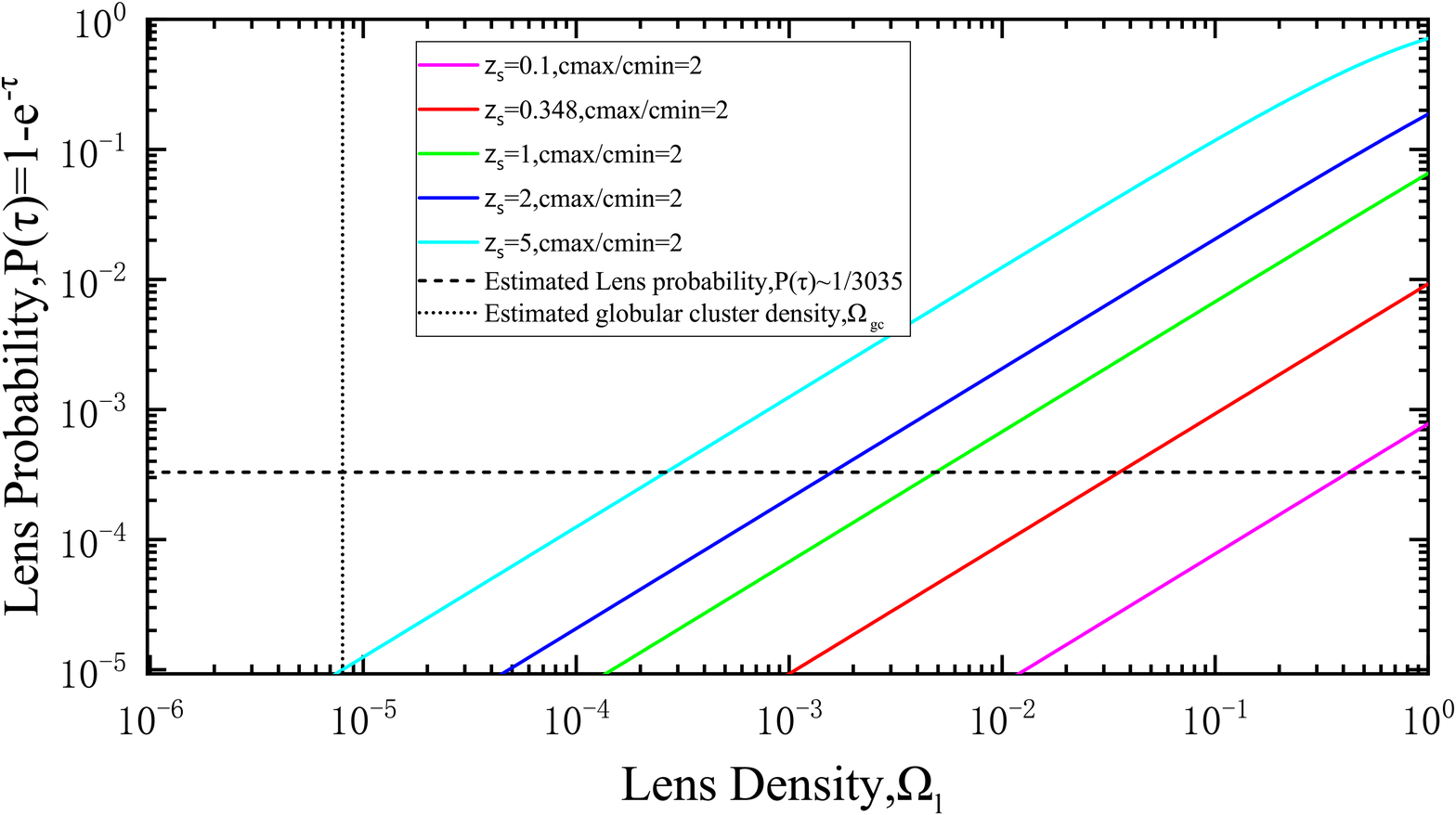}
\caption{Optical depth as a function of source redshift $z_s$ with a $C_{\rm max}/C_{\rm min}$ value of 2.0.
Different color lines correspond to different values of $z_s$. The dashed black horizontal line is the estimated lens probability based
on seeing one event in 3035 light curves. The dotted black vertical line is the estimated globular cluster density $\Omega_{\rm gc}$.}
\label{fig:Number-density}
\end{figure}

\end{document}